\documentclass[useAMS,usenatbib]{mn2e}
\usepackage{float} 
\usepackage{graphicx}
\usepackage{textcomp}
\usepackage{natbib}
\usepackage{lscape}
\usepackage{deluxetable}
\usepackage{mathrsfs}

\usepackage{amsmath} 
\usepackage{amssymb}
\newcommand{\um}{\textmu m }
\newcommand{\uu}{\textmu m}

\newcommand{\ssil}{$S\!_{\rm{sil}}$ }
\newcommand{\ssilu}{$S\!_{\rm{sil}}$}

\newcommand{\lum}[1]{$L_{\rm{#1}}$} 
\newcommand{\nev}{[Ne {\sc v}] }
\newcommand{\neiii}{[Ne {\sc iii}] }
\newcommand{\neii}{[Ne {\sc ii}] }

\newcommand{\arii}{[Ar {\sc ii}] }

\newcommand{\siv}{[S {\sc iv}] }
\newcommand{\clii}{[Cl {\sc ii}] }

\newcommand{\ewratio}{$EW_{\rm{12.7}}$/$EW_{\rm{11.2}}$}
\newcommand{\pahratio}{$f_{\rm{12.7}}$/$f_{\rm{11.2}}$}
\newcommand{\od}[1]{$\tau_{\rm{#1}}$} 

\newcommand{\ew}[1]{$EW_{\rm{#1}}$}

\bibpunct[]{(}{)}{;}{a}{}{,}

\title[Extinction in the 11.2 \um PAH band]{Extinction in the 11.2 \um PAH band and the low \lum{11.2}/\lum{IR} in ULIRGs}

\author[A. Hern\'an-Caballero et al.]{
Antonio Hern\'an-Caballero,$^{1}$\thanks{E-mail: ahernan@cefca.es} Henrik W. W. Spoon,$^2$
Almudena Alonso-Herrero,$^3$\newauthor
Evanthia Hatziminaoglou,$^4$
Georgios E. Magdis,$^{5,6,7}$
Pablo G. P\'erez-Gonz\'alez,$^8$\newauthor
Miguel Pereira-Santaella,$^8$
Santiago Arribas,$^8$
Isabella Cortzen,$^{5,7}$
\'Alvaro Labiano,$^3$\newauthor
Javier Piqueras,$^8$
and Dimitra Rigopoulou$^9$\\
$^{1}$Centro de Estudios de F\'isica del Cosmos de Arag\'on (CEFCA), Plaza San Juan, 1, E-44001 Teruel, Spain\\
$^{2}$Cornell Center for Astrophysics and Planetary Science, Ithaca, NY 14853, USA\\
$^{3}$Centro de Astrobiolog\'ia, (CAB, CSIC-INTA), ESAC Campus, E-28692 Villanueva de la Ca\~nada, Madrid, Spain\\
$^{4}$European Southern Observatory, Karl-Schwarzschild-Str. 2, 85748 Garching bei M\"unchen, Germany\\
$^{5}$Cosmic Dawn Center (DAWN), Copenhagen, Denmark\\
$^{6}$DTU-Space, Technical University of Denmark, Elektrovej 327, DK-2800 Kgs. Lyngby, Denmark\\
$^{7}$Niels Bohr Institute, University of Copenhagen, Lyngbyvej 2, DK-2100 Copenhagen {{\O}}, Denmark\\
$^{8}$Centro de Astrobiolog\'ia, (CAB, CSIC-INTA), Carretera de Ajalvir, km 4, E-28850 Torrej\'on de Ardoz, Madrid, Spain\\
$^{9}$Department of Physics, University of Oxford, Keble
Road, Oxford OX1 3RH, UK\\}

\begin{document}
\date{Accepted ........ Received ........;}

\pagerange{\pageref{firstpage}--\pageref{lastpage}} \pubyear{2019}

\maketitle

\label{firstpage}

\begin{abstract}
We present a method for recovering the intrinsic (extinction-corrected) luminosity of the 11.2 \um PAH band in galaxy spectra. Using 105 high S/N \textit{Spitzer}/IRS spectra of star-forming galaxies, we show that the equivalent width ratio of the 12.7 \um and 11.2 \um PAH bands is independent on the optical depth ($\tau$), with small dispersion ($\sim$5\%) indicative of a nearly constant intrinsic flux ratio $R_{int}$ = (\pahratio)$_{int}$=0.377 $\pm$ 0.020.
Conversely, the observed flux ratio, $R_{obs}$ = (\pahratio)$_{obs}$, strongly correlates with the silicate strength (\ssilu) confirming that differences in $R_{obs}$ reflect variation in $\tau$.
The relation between $R_{obs}$ and \ssil reproduces predictions for the Galactic Centre extinction law but disagrees with other laws. 
We calibrate the total extinction affecting the 11.2 \um PAH from $R_{obs}$, which we apply to another sample of 215 galaxies with accurate measurements of the total infrared luminosity (\lum{IR}) to investigate the impact of extinction on \lum{11.2}/\lum{IR}.
Correlation between \lum{11.2}/\lum{IR}{} and $R_{obs}$ independently on \lum{IR} suggests that increased extinction explains the well known decrease in the average \lum{11.2}/\lum{IR} at high \lum{IR}. The extinction-corrected \lum{11.2} is proportional to \lum{IR}{} in the range \lum{IR}{} = 10$^9$--10$^{13}$ L$_\odot$. These results consolidate \lum{11.2} as a robust tracer of star formation in galaxies.

\end{abstract}


\section{Introduction} 

The mid-infrared (MIR) spectrum of star-forming galaxies is characterised by strong emission bands from polycyclic aromatic hydrocarbon (PAH) molecules that populate the interstellar medium (ISM).  
Both theoretical and observational studies point to a large diversity of chemical species, including neutral, ionised, and substituted or complexed molecules of varying sizes \citep[see e.g. the review by][ and references therein]{Tielens08}.
The PAH bands arise from fluorescence of excited PAH molecules, pumped by ultraviolet and optical photons \citep{Allamandola89}. Unlike atomic transitions or molecular lines, individual PAH bands are not associated to a single species, but to specific stretching and bending modes of the C-H and C-C links, which take slightly different energies in each species. Accordingly, the profile and strength of an individual PAH band depends on the relative abundance of PAH molecules with different sizes or spatial structure \citep[e.g.][]{Hudgins05,Peeters11,Sloan14}.

The predominance of PAH emission in the MIR spectra of regions of massive star formation (SF) makes the PAH bands useful indicators of the star formation rate (SFR). The lower opacity of the interstellar dust at MIR wavelengths compared to the UV, optical, and near-infrared (NIR) implies that the PAH bands can trace obscured SF that is undetected in the UV, H$\alpha$ or even Pa$\alpha$ observations \citep[e.g.][]{Rieke09}, while the high luminosity of the strongest PAH features makes them competitive with other SFR tracers at high redshifts. 
The PAH emission is weak inside HII regions, with most of the PAH flux arising from the photodissociation regions (PDRs) that surround them and the diffuse ISM. In spite of this, a tight correlation is found between the integrated luminosity of the PAH bands, \lum{PAH}, and the SFR at galaxy-wide and sub-kpc scales in systems covering a broad range of stellar masses, redshifts, and SFR \citep[e.g.][]{Farrah07,Sargsyan09,Treyer10,Diamond-Stanic12,Shipley16} although with a dependence on the metallicity \citep{Calzetti07}. 
 
Two strong PAH bands at 6.2 \um and 11.2 \um have been extensively used as SFR indicators. Their main advantage over other PAH features is their relative isolation, with narrow and nearly constant profiles (unlike the much broader and variable complexes at $\sim$7.7 \um and $\sim$17 \uu), that facilitate their measurement.
While the 6.2 \um band captured the most interest at first, it presents some important drawbacks: the 6.2 \um PAH resides inside a broad water ice absorption band peaking at $\sim$6.0 \uu, which is often not discernible due to infilling by the PAH band. This water ice band depresses the continuum underlying the PAH band, causing an underestimation of the actual PAH flux if not accounted for. In the case of a high dust column, as is the case in most ultra-luminous infrared galaxies (ULIRGs), the 6.2 \um PAH may be strongly depleted or even completely suppressed by this effect \citep[see e.g.][]{Spoon06}. While the 11.2 \um PAH is also located within an absorption band (from silicates, peaking at $\sim$9.8\uu), the much wider profile of the silicate band facilitates the interpolation of the continuum underlying the 11.2 \um PAH feature.
Another disadvantage of the 6.2 \um band (and also those at 7.7 \um and 8.6 \uu) is that they might be suppressed relative to 11.2 \um in the vicinity of AGN due to destruction of the smaller molecules by shocks or by the X-ray emission of the AGN \citep{Smith07,Wu10}, while the 11.2 \um band remains a good tracer of the SFR in AGN host galaxies \citep{Diamond-Stanic10,Esquej14}. Lastly, the 11.2 \um PAH is the only strong PAH band at a wavelength observable from the ground in local galaxies, meaning that only the 11.2 \um PAH can trace the SFR on sub-kpc scales (with 8--10m-class telescopes) until the \textit{James Webb Space Telescope} (\textit{JWST}) becomes available. 

The importance of the 11.2 \um PAH as a SF tracer in the circumnuclear region of AGN cannot be overstated. Traditional SFR indicators (such as the UV continuum, hydrogen recombination lines, and the \neii 12.81 \um line) suffer from contamination by the AGN emission, while measurements based on the far-infrared luminosity suffer from very poor spatial resolution \citep[e.g.][]{Garcia-Gonzalez16}. MIR spectroscopy of local Seyferts with sub-arcsecond resolution shows 11.2 \um PAH emission at distances of a few tens of pc from the AGN \citep{Honig10,Gonzalez-Martin13,Alonso-Herrero14,Alonso-Herrero16} and the 11.2 \um PAH luminosity indicates a correlation between the (circum-)nuclear SFR and the accretion rate onto the supermassive black hole \citep{Esquej14,Ruschel-Dutra17,Esparza-Arredondo18}, although \citet{Jensen17} caution that the AGN might also contribute to excite the PAH emission in the central kpc. 

Confidence in the PAH luminosity (including the 11.2 \um band) as a first-rate SFR indicator is undermined by its perceived dependence on the surface density of star formation.
In the local Universe, there is evidence for a decreasing contribution from PAHs to the total infrared luminosity (\lum{IR}) with increasing \lum{IR}, in particular in the ULIRG (\lum{IR} $>$ 10$^{12}$ L$_\odot$) regime \citep[e.g.][]{Rigby08,Rieke09}. On the contrary, high redshift ULIRGs show extended SF on kpc scales and \lum{PAH}/\lum{IR}{} ratios comparable to those of lower luminosity local galaxies \citep[e.g.][]{Farrah08,Muzzin10,Elbaz11}. 
This has been interpreted as a consequence of the higher SFR density in local ULIRGs, which have most of their SF concentrated within a compact core. A higher density of SF could suppress PAH emission by destroying or ionising the PAH carriers with a stronger radiation field. This interpretation is supported by the increased 6.2/11.2 PAH ratios in local ULIRGs relative to both lower luminosity starbursts and high redshift ULIRGs \citep[e.g.][]{Farrah08,Hernan-Caballero09}, since the 6.2/11.2 ratio correlates with the ionised fraction of PAH carriers \citep{Boersma16}.

Alternatively, low \lum{PAH}/\lum{IR}{} could be just a geometrical effect of concentrating the SF in a few compact regions, which reduces the PAH emitting surface from PDRs relative to the same amount of SF distributed in more PDRs. This is evidenced by the systematic differences in the rest-frame 8 \um to IR luminosity ratio, \lum{8}/\lum{IR}, between galaxies in the `normal' and `starburst' modes of star formation \citep{Elbaz11,Elbaz18,Magdis13}.
Another interpretation of the lower \lum{PAH}/\lum{IR}{} in local ULIRGs that does not require PAH suppression claims that \lum{PAH} ultimately traces the molecular gas content, not star formation \citep{Cortzen19}. They base their hypothesis in their finding of a stronger correlation between the 6.2 \um PAH luminosity (\lum{6.2}) and the CO luminosity (\lum{CO}) compared to that between \lum{6.2}{} and \lum{IR}. 

All these interpretations fail to acknowledge the impact of dust obscuration on the observed PAH luminosity. While the extinction at MIR wavelengths is less than in the optical or NIR, it is still significant (the extinction at the peak of the $\sim$9.8 \um silicate feature is comparable to that of Pa$\alpha$). This means that underestimation of the intrinsic PAH luminosity due to dust obscuration may contribute to the lower \lum{PAH}/\lum{IR}{} found in local ULIRGs and also adds dispersion to the correlation between \lum{PAH}{} and \lum{IR}.

The main reason why dust obscuration is often overlooked in PAH luminosity measurements is the difficulty in obtaining realistic estimates of the amount of extinction affecting the PAH bands.  
If the extinction law is known, the extinction at the wavelength of the PAH bands can in principle be inferred from that measured through the hydrogen recombination lines or the MIR H$_2$ rotational lines. However, extinction measurements from optical recombination lines are biased against the most obscured regions, while NIR and MIR lines (the Paschen, Bracket, and Pfund series), as well as the $H_2$ lines, are usually too faint to be useful beyond the local Universe. In addition, by using these lines to estimate the extinction in the PAH bands we implicitly assume that the dust column density towards the PAH emitting regions is the same as that of the recombination or H$_2$ lines, which may not always be the case given that they originate in different environments (PDRs and the diffuse ISM for PAHs, HII regions for recombination lines, and warm molecular gas for H$_2$). 

Another method for quantifying the extinction towards star-forming regions is to measure the optical depth of the 9.8 \um silicate absorption band \citep{Roche84,Whittet03}. It is usually estimated from the `silicate strength' \citep[\ssilu;][]{Spoon06} which measures the depth of the 9.8 \um silicate band relative to an interpolated continuum. Unlike extinction estimates from emission lines, \ssil does not require high spectral resolution or high S/N data, and it has been extensively used in the literature \citep[e.g.][]{Hao07,Hatziminaoglou15}. However, since it relies on a single absorption band, it is affected by infilling from any additional continuum emission. In particular, the strong continuum emission from the AGN in composite sources may increase or decrease the depth of the silicate band (or even make it appear in emission) depending on the spectrum and relative luminosity of the AGN and host galaxy components \citep[see e.g.][]{Hernan-Caballero15}.

As a consequence of all these complications, usually no attempts are made at correcting PAH luminosities for extinction. Instead, calibrations of the SFR as a function of the apparent PAH luminosity implicitly assume an effective $\tau$ that is some average of the values for the individual galaxies used in the calibration. This implies that galaxies with higher (lower) than usual extinction get their SFR under- (over-)estimated. Neglecting the extinction therefore adds dispersion to the relation between the PAH luminosity and other SFR indicators, and biases SFR estimates in dusty galaxies. 
 
In this work we demonstrate a novel approach to determine the total extinction affecting the 11.2 \um PAH band, $A_{11.2}$, that overcomes all the issues presented above. The method takes advantage of differential extinction between the 11.2 \um and 12.7 \um PAH bands as well as a very small dispersion in their \textit{intrinsic} flux ratio on galaxy scales to determine the extinction at 11.2 \um from the \textit{observed} flux ratio. 
The structure of the paper is as follows: in \S2 we describe the two samples of star-forming galaxies used, respectively, to calibrate the relation between the PAH flux ratio and extinction and to check how the extinction-correction of the 11.2 \um PAH luminosity affects the correlation with \lum{IR}. In \S3 we discuss in length the physics of the 12.7/11.2 PAH ratio and the causes for its variation, and use the empirical relation between the 12.7/11.2 PAH ratio and the depth of the silicate feature to estimate the intrinsic value of the 12.7/11.2 PAH ratio and identify the extinction law that best reproduces the extinction in star-forming regions. In \S4 we apply our extinction correction to the sample with accurate far infrared measurements of \lum{IR} and discuss the implications for the luminosity dependence of the \lum{11.2}/\lum{IR} relation. Finally, \S5 outlines some prospects for future observations with \textit{JWST} and \textit{SPICA}, and \S6 summarises our main conclusions.

\section{Data}
\subsection{The IDEOS database}

Our parent sample is the set of 3532 galaxies in the Infrared Database of Extragalactic Observables from \textit{Spitzer} (IDEOS)\footnote{http://ideos.astro.cornell.edu}. 
IDEOS provides homogeneously measured spectroscopic observables for nearly all the galaxies beyond the Local Group that were observed with the Infrared Spectrograph \citep[IRS;][]{Houck04} onboard \textit{Spitzer} using the low-resolution $R\sim$ 60--120 modules.\footnote{IDEOS includes all the spectra of galaxies obtained in the staring mode. Nearby galaxies observed in the spectral mapping mode are excluded.}
The stitching of the different spectral orders in the \textit{Spitzer}/IRS spectra, cross-identification with the NASA Extragalactic Database (NED), and determination and validation of redshifts is presented in Hern\'an-Caballero et al. (2016). 

From the IDEOS database we extract measurements of redshift, \ssilu, PAH fluxes and equivalent widths (EW), and fluxes for neon lines (\neii 12.81 \uu, \neiii 15.56 \uu, \nev 14.32 \uu, and \nev 24.32 \uu). A thorough description of the methods employed in these measurements is presented in Spoon et al. (in preparation). A brief summary follows. 

Fluxes and equivalent widths for the 6.2, 11.2, and 12.7 \um PAH bands are measured by fitting the 5.39--7.35 \um and 9.8--13.5 \um spectral ranges with a combination of: polynomial continuum, Pearson type-IV distribution profiles \citep{Pearson95} for the 6.2, 11.2, and 12.7 \um PAH bands, Gaussian profiles for the much fainter PAH bands at 5.68, 6.04, 10.64, 11.04, and 12.00 \uu, and also gaussian profiles for the (unresolved) emission lines of H$_2$ (5.51, 6.91, and 12.28 \uu), \arii (6.99 \uu), \siv (10.51 \uu), and \neii (12.81 \uu).

Fluxes of the \nev 14.32 \um and \neiii 15.56 \um lines are obtained by fitting Gaussian profiles on top of a 4$^{th}$ order polynomial continuum. The \nev 14.32 \um line is blended with a \clii{} line at 14.37 \um and a PAH band at 14.22 \uu, which are simultaneously fitted with Gaussian profiles.

The silicate strength is defined \citep[e.g.][]{Spoon07} as: \ssil = $\ln$ [$f^{obs}$($\lambda_p$)/$f^{cont}$($\lambda_p$)], where $\lambda_p$ is the wavelength of the peak of the silicate feature (usually $\lambda_p$$\sim$9.8 \um when found in absorption), $f^{obs}$($\lambda_p$) is the flux density measured at $\lambda_p$, and $f^{cont}$($\lambda_p$) is the flux density of the underlying continuum that would be measured at $\lambda_p$ in absence of silicate emission/absorption. The latter is usually estimated by interpolation of the spectrum between anchor points at wavelengths outside of the range covered by the silicate feature, where the opacity is much lower. \ssil takes positive (negative) values when the silicate feature appears in emission (absorption).

Measurements of \ssil in IDEOS assume either a spline or power-law shape for the underlying continuum. For sources with strong PAH emission like the ones in our samples, power-law interpolation between anchor points at 5.5 \um and 14 \um is preferred. The flux at the 9.8 \um peak of the silicate feature is measured by fitting the silicate profile with a 4$^{th}$ order polynomial after removing the PAH bands and emission lines fitted in the previous step. 

We obtained morphological and spectroscopic classifications of the IDEOS sources from NED. We group spectroscopic classifications in three broad categories: AGN (includes Seyfert 1/1.x/2, QSOs, LINERs, radio galaxies, Blazars, `candidate' AGN, and AGN with uncertain classifications), starbursts (includes starburst and HII galaxies), and other (including other non-AGN classifications or no classification at all).

\subsection{Selection of the calibration sample}

The calibration sample is a subsample of IDEOS carefully selected to provide an accurate measurement of the intrinsic (extinction-corrected) value of \pahratio{} and its dispersion in star-forming galaxies, and to obtain an empirical relation between \pahratio{} and \od{11.2}{} that can be applied to correct for extinction the 11.2 \um PAH luminosity in any galaxy. 
We require full spectral coverage in the rest-frame range 5.5--16 \um in order to obtain all the necessary MIR diagnostics, and detection of the 11.2 \um and 12.7 \um PAH bands with S/N greater than 40 and 15, respectively, to ensure that flux and EW ratios have uncertainties of less than 10\%. 

From 261 IDEOS sources meeting these requirements, we exclude 113 sources with spectroscopic classification as AGN following the criteria described above.
Because some sources have no spectroscopic classification in NED, and others may host obscured AGN that do not show up in the optical, we further clean the sample by removing sources meeting any of these additional criteria suggestive of the presence of an AGN: 
a) 9.8 \um silicate feature in emission (\ssil $>$ 0);
b) Low equivalent width of the 6.2 \um PAH (\ew{6.2}{} $<$ 0.8 \uu);
c) detection at $>$2$\sigma$ of the \nev 14.32 \um or \nev 24.32 \um lines.

Finally, we also exclude a few inactive galaxies morphologically classified as ellipticals, which are known to often show abnormal PAH ratios \citep{Kaneda08}, as well as galaxies with \neiii/\neii $>$ 0.7, suggestive of a stronger than usual interstellar radiation field that may also influence PAH ratios \citep{Hunt10}. The resulting sample contains 105 sources at redshift $z$$<$0.12.

\subsection{Selection of the far infrared sample}\label{sec:firsample}

The FIR sample is intended to test how the extinction correction influences the relation between the 11.2 \um PAH luminosity, \lum{11.2}, and the total infrared luminosity, \lum{IR}. 
We draw this sample from the 5 Milli-Jansky Unbiased Spitzer Extragalactic Survey \citep[5MUSES;][]{Wu10}, and the Herschel Multi-tiered Extragalactic Survey \citep[HerMES;][]{Oliver12}. 
The 5MUSES sample combines \textit{Spitzer}/IRS spectroscopy and far-infrared (FIR) photometry from both \textit{Spitzer}/MIPS (24, 70, 160 \uu) and \textit{Herschel}/SPIRE (250, 350, 500 \uu), while HerMES provides photometry in the 3 SPIRE bands plus the \textit{Herschel}/PACS bands at 100 \um and 160 \um in some fields. We take the HerMES photometry for \textit{Spitzer}/IRS sources from the matched HerMES-IRS catalog presented by \citet{Feltre13}.

From both samples we select sources with S/N $>$ 4-$\sigma$ in the 12.7 \um PAH band and the 250 \um flux density. 
We exclude a few sources with uncertain redshifts \citep[Quality flag $<$ 3; see][ for details]{Hernan-Caballero16} or insufficient spectral coverage to measure the 11.2 \um or 12.7 \um PAH bands.
The \lum{IR}{} was obtained by fitting the available photometry in the observed 24--500 \um range with dust emission models from \citet{Draine07} and integrating the 8--1000 \um rest-frame range \citep[see ][ for details]{Magdis13}. 
Our final FIR sample contains 215 sources at redshifts between $z$=0.022 and $z$=1.71 (mean: 0.287, median: 0.143), four of them also included in the calibration sample.

\section{The 12.7/11.2 PAH ratio}\label{sec:PAHratio}

\begin{figure} 
\begin{center}
\includegraphics[width=8.4cm]{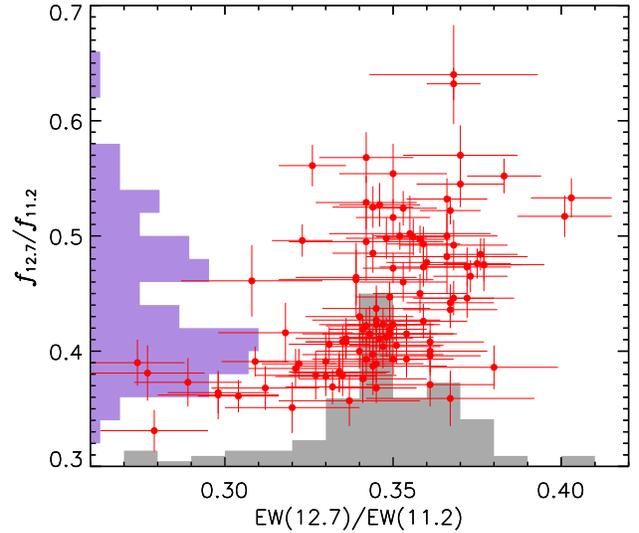}
\end{center}
\caption[]{Distribution of the ratio between measured fluxes in the PAH bands at 12.7 \um and 11.2 \uu, \pahratio, and the ratio of their respective equivalent widths, \ewratio, for the 105 sources in the calibration sample.\label{fig:EWPAHratio}}
\end{figure} 

\subsection{Physics of the 12.7/11.2 PAH ratio}

PAH bands in the 10--15 \um range are generally associated with out-of-plane bending vibrational modes of hydrogen atoms \citep[e.g.][]{Hony01}. The wavelength of the emission is determined by the number of adjacent C--H groups, with the 11.2 \um feature arising from solo C-H groups while the 12.7 \um feature is attributed to trio groups \citep{Hony01,Bauschlicher08}.

Galactic HII regions and reflection nebulae show significant variation both spatially and among sources in the flux ratio between the 12.7 \um and 11.2 \um bands, \pahratio. This was first attributed to dehydrogenation \citep{Duley81}, and later to changes in the molecular structure and/or erosion \citep{Hony01,Fleming10,Boersma12}.
However, laboratory studies highlight the importance of the charge state of PAH molecules in determining the relative intensities of the different PAH bands \citep{Allamandola99,Galliano08}. 

Because the intensity of the 11.2 \um band correlates with the one at 3.3 \uu, and the latter is very weak in ionised PAHs \citep{Langhoff96}, it has been widely assumed that the dominant component in the 11.2 \um band arises from neutral PAHs \citep[e.g.][]{Allamandola99,Li01, Hony01,Smith07}. However, a contribution from cations is required to interpret the subtle source-to-source variation in the peak wavelength and shape of the profile \citep{Boersma16,Shannon16}. In the case of the 12.7 \um band the situation is more complex, with several blended components from cations and neutrals contributing to the observed profile \citep[e.g.][]{Shannon16}.
Accordingly, \pahratio{} should be sensitive to the relative fraction of cations in the population of PAH molecules. Indeed, a correlation is found between the PAH ionisation parameter, $G_0 T^{1/2}_{gas}/n_e$ (where $G_0$ is the strength of the local radiation field, $T_{gas}$ is the temperature of the gas, and $n_e$ the electron density), and the 12.7/(11.2+12.7) flux ratio \citep[e.g.][]{Galliano08,Fleming10,Boersma16}.
A comprehensive analysis of the 11.2 \um and 12.7 \um PAH bands in spectral maps of two reflection nebulae and a star-forming region by \citet{Shannon16} concluded that PAH charge is indeed the main factor driving variation in \pahratio, while molecular structure plays a secondary role. 

\subsection{The effect of extinction}

\begin{figure} 
\begin{center}
\includegraphics[width=8.4cm]{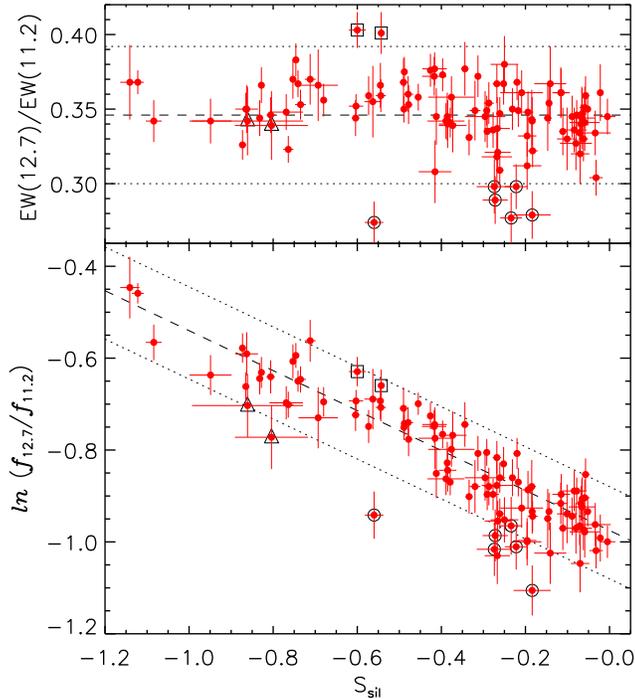}
\end{center}
\caption[]{(Top panel) Dependency of the ratio of the equivalent widths of the 12.7 \um and 11.2 \um PAH bands with the silicate strength. The dashed line indicates the average \ewratio{} for the sample, while the dotted lines enclose the $\pm$2$\sigma$ interval. (Bottom panel) observed flux ratio between the two PAH bands as a function of \ssilu. The dashed line represents a weighted linear fit to the sample, while the dotted lines mark a $\pm$10\% deviation from this fit. Open squares, triangles, and circles highlight sources with unusual \ewratio, \pahratio, or both (see text).\label{fig:PAH-vs-Ssil}}
\end{figure} 

\citet{Shannon16} acknowledge the impact of extinction in \pahratio{} by pointing out that the largest values are found in more obscured regions, and that applying an extinction correction greatly reduces the range of \pahratio{}.
This is because the dust opacity at 11.2 \um is high relative to 12.7 \uu, although with a strong dependence on the choice of the extinction law (see \S\ref{sec:extlaw}).
An earlier hint on the important role of extinction was the discovery by \citet{Hony01} of an anti-correlation between \pahratio{} and the relative contribution of the PAHs to the total infrared luminosity, \lum{PAH}/\lum{IR}. While they interpreted this as evidence for sensitivity of \pahratio{} to the size distribution of PAH molecules, it is also consistent with the effect of extinction, which increases \pahratio{} and decreases \lum{PAH}/\lum{IR}.
Therefore, it is likely that the observed dispersion in \pahratio{} found in Galactic HII regions and reflection nebulae is a consequence of spatial variations in both the local ionisation parameter and the extinction along the line of sight.

In \textit{Spitzer}/IRS observations of extragalactic sources the spatial resolution element encompasses entire star formation regions or even whole galaxies, which has the effect of averaging out the spatial variation in the ionisation parameter. As a consequence, extinction might become the main factor driving the observed dispersion in \pahratio{} for individual galaxies.
To test this hypothesis, we rely on the fact that equivalent widths are not affected by extinction (because both the PAH band and the underlying continuum are suppressed in the same way). Therefore, source to source variation in the ratio between the EW of the 12.7 \um and 11.2 \um PAH bands, \ewratio, may depend on systematic differences in the molecular structure and charge of the PAH molecules (e. g. due to different gas metallicity), but not on differences in the amount of extinction.
Accordingly, the dispersion in \ewratio{} quantifies the intrinsic dispersion in \pahratio{} that would be observed in absence of extinction. 
    
Figure \ref{fig:EWPAHratio} shows the distribution of \ewratio{} and \pahratio{} for the 105 sources in our calibration sample. 
The average value of \ewratio{} is $\langle$\ewratio$\rangle$ = 0.346, and the 1-$\sigma$ dispersion is $\sigma$(\ewratio) = 0.023. The average 1-$\sigma$ uncertainty in \ewratio{} measurements for individual sources is 0.015, which subtracted in quadrature implies that the intrinsic dispersion in \ewratio{} for the population is $\sim$0.017, or $\sim$5\% of the average value. 
The observed dispersion in \pahratio{} for our sample is much higher, at 14.3\% of the average value ($\langle$\pahratio$\rangle$ = 0.443, $\sigma$(\pahratio) = 0.064), with a negligible contribution from photometric uncertainty (the average 1-$\sigma$ error in \pahratio{} for individual sources is 0.018).
The fact that the observed dispersion in \pahratio{} is much larger than in \ewratio{} suggests that extinction is the dominant factor in producing the former. 

It is worth noting that any alternative mechanism for explaining the observed dispersion in \pahratio{} would need to carefully adjust its impact on the PAH flux and the underlying continuum so that the EW remains unchanged. 
Further evidence on the dominant role that extinction plays in the observed \pahratio{} is its strong correlation with the strength of the silicate absorption feature, \ssil (Figure \ref{fig:PAH-vs-Ssil}, bottom panel). 
In Section \ref{sec:extlaw} we will use this to calibrate the relation between \pahratio{} and the dust column density and to put constraints on the shape of the extinction law. 

\begin{figure} 
\begin{center}
\includegraphics[width=8.4cm]{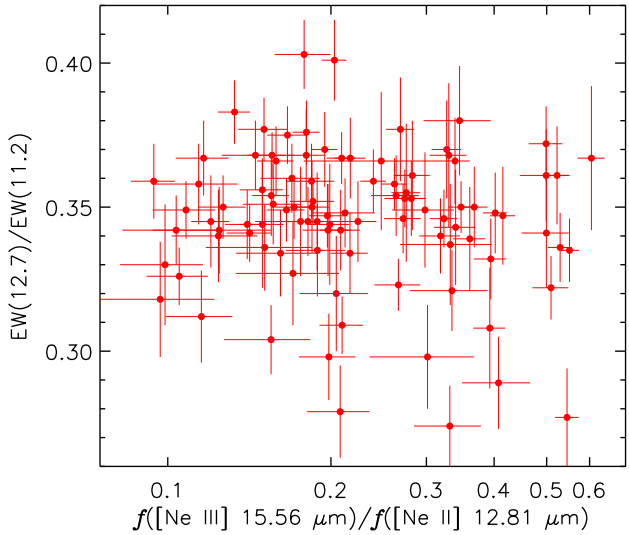}
\end{center}
\caption[]{Ratio of PAH equivalent widths as a function of the \neiii/\neii line intensity ratio for the 97 sources with detection of both lines in the calibration sample.\label{fig:Neon}}
\end{figure} 

\subsection{The intrinsic 12.7/11.2 flux ratio}\label{sec:intrinsic-ratio}

The top panel in Figure \ref{fig:PAH-vs-Ssil} shows that \ewratio{} has no significant dependence with \ssilu. This is important because it indicates that the intrinsic \pahratio{} does not change from lightly to heavily obscured sources. Figure \ref{fig:Neon} shows that \ewratio{} is also independent on the \neiii/\neii ratio (correlation coefficient $r$ = -0.13). This suggests significant resilience to variation in the hardness of the radiation field, at least for the range covered by our calibration sample (\neiii/\neii $<$ 0.7), which includes $\sim$90\% of HII galaxies \citep[see Figure 9 in][]{Pereira-Santaella10}.
Combined with the fact that photometric errors account for most of the dispersion in \ewratio, it becomes apparent that for practical purposes the intrinsic value of the \pahratio{} ratio, $R_{int}$, may be considered constant for the population of star-forming galaxies represented by our calibration sample. 

In Appendix \ref{appendixb} we demonstrate that the observed value of the \pahratio{} ratio, $R_{obs}$, is related to \ssil and $R_{int}$ by:

\begin{equation}
\ln R_{obs} = \ln R_{int} + \xi S_{sil}
\end{equation}

\noindent where the parameter $\xi$ is a constant that only depends on the shape of the extinction law and the anchor points used to define \ssil (see \S\ref{sec:extlaw}).

We estimate the values of $\xi$ and $R_{int}$ by fitting a linear relation between $\ln$($R_{obs}$) and \ssil (Figure \ref{fig:PAH-vs-Ssil}, bottom panel). We obtain $\xi$ = -0.435 $\pm$ 0.013 and $R_{int}$ = 0.377 $\pm$ 0.003 (statistic) $\pm$ 0.016 (systematic), where the latter term arises from a systematic error of $\sim$0.1 in \ssil (Spoon et al. in preparation) which is caused by the uncertainty in the determination of the intrinsic continuum at 9.8 \uu, interpolated from anchor points at 5.5 \um and 14 \um (see appendix \ref{appendixb}).

\begin{figure} 
\begin{center}
\includegraphics[width=8.4cm]{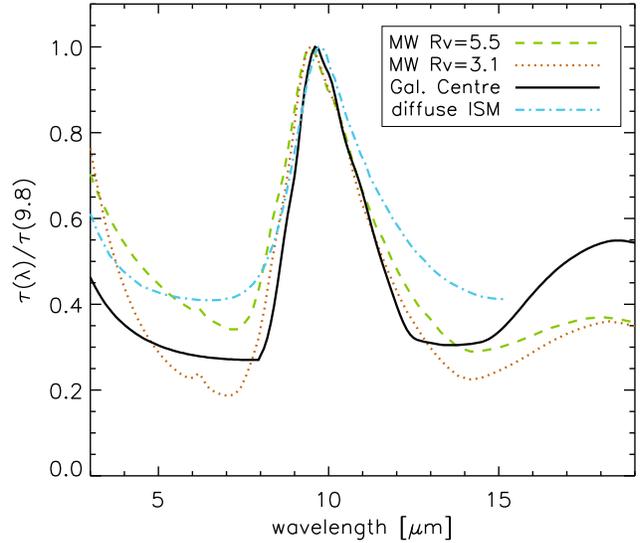}
\end{center}
\caption[]{Representative examples of the variety of mid-IR extinction laws in the literature, all normalised at their peak opacity near $\sim$9.8 \um (see text for details).\label{fig:extlaws}}
\end{figure} 

Individual sources that deviate significantly from this linear relation may do it for several reasons. Outliers in both \pahratio{} and \ewratio{} (sources highlighted with big circles in Figure \ref{fig:EWPAHratio}) are likely to have a different $R_{int}$. If the source is far-off in the EW ratio but not the PAH ratio (open squares), it might be that the continuum underlying one of the PAHs is under- or over-estimated. Finally, if the separation occurs only in the PAH flux ratio (open triangles), the most likely interpretation is that \ssil does not represent accurately the obscuration towards the PAH emission in that particular source. This may indicate a problem in the \ssil measurement or a small contamination of the MIR continuum by an unidentified AGN. 

After removing this $\sim$10\% of problematic sources, the difference $\Delta$\ssil between the \ssil measured on the spectrum and the one predicted from 
$R_{obs}$ using the linear relation has a dispersion $\sigma$($\Delta$\ssilu) = 0.11, comparable to the systematic uncertainty of individual \ssil measurements.
This means that for this sample of star-forming galaxies $R_{obs}$ traces the optical depth with an accuracy comparable to what can be obtained from the silicate feature. However, unlike \ssilu, $R_{obs}$ is a valid tracer of the optical depth towards star-forming regions also for composite sources, where the
emission from AGN-heated dust modifies the depth of the silicate feature in the integrated spectrum \citep[see][]{Hernan-Caballero15}, while the PAH fluxes are not affected. 

\subsection{Constraints on the extinction law}\label{sec:extlaw}

Laboratory measurements and observations of astrophysical sources show that the extinction law of the interstellar dust is dominated at MIR wavelengths by the absorption cross-section of silicate grains, which produces two broad peaks in opacity centred at $\sim$9.8 \um and $\sim$18 \uu. In between the two peaks, the opacity decreases reaching a minimum at $\sim$14 \uu. The shape of the $\sim$9.8 \um peak (which encompasses the PAH bands at 11.2 \um and 12.7 \uu) depends on the chemical and mineralogical composition of the silicate grains, with different dust models and empirical measurements in a range of environments producing wildly different profiles. 

Figure \ref{fig:extlaws} shows the opacity profiles around the $\sim$9.8 \um peak of the silicate absorption for some representative examples of extinction laws from the literature: a model for the Milky Way extinction assuming R$_V$ = 5.5 \citep{Weingartner01}, a Milky Way model with R$_V$ = 3.1 and 60 ppm of C in PAHs \citep{Draine03}, the ``Galactic Centre'' extinction law from the absorption profile of the highly obscured Wolf Rayet star GCS-3 \citep{Chiar06}, and the extinction law of the diffuse ISM obtained from the average absorption profile of five OB stars \citep{Shao18}. 

We can use the slope $\xi$ of the relation between $\ln$($R_{obs}$) and \ssil calculated in \S\ref{sec:intrinsic-ratio} to identify the extinction law that best reproduces the shape of the silicate profile for the galaxies in our calibration sample.
We compute $\xi$ for the extinction laws shown in Figure \ref{fig:extlaws} assuming two idealised cases for the dust geometry: a screen model (where all the obscuring dust is in the foreground) and a mixed model (where dust and PAH-emitting molecules are homogeneously mixed). In real galaxies, the actual value is likely to be intermediate between the screen and mixed models, with some extinction due to dust in the star-forming regions that produce the PAH emission and some additional extinction in the foreground.

\begin{figure} 
\begin{center}
\includegraphics[width=8.4cm]{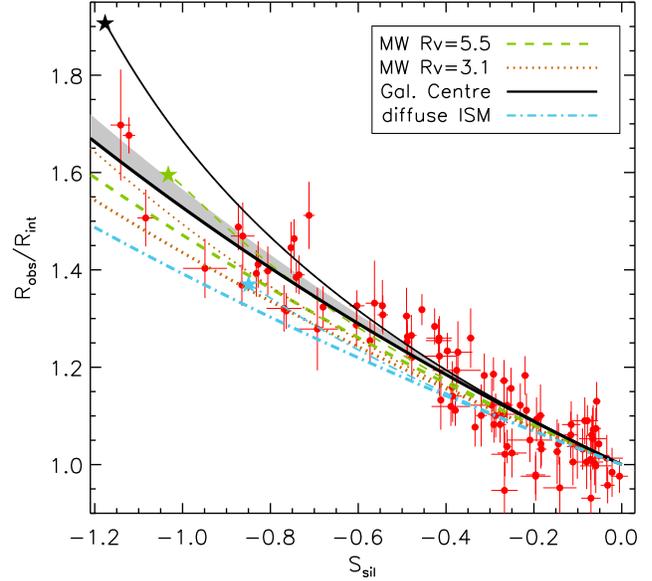}
\end{center}
\caption[]{Predicted relation between the observed and intrinsic flux ratios of the 12.7 \um and 11.2 \um PAH bands as a function of the silicate strength for the extinction laws in Figure \ref{fig:extlaws}. Thick (thin) lines represent the prediction assuming a screen (mixed) dust geometry. Stars indicate the limits corresponding to $\tau \rightarrow \infty$ for the mixed model (see text). The solid grey area corresponds to the 1-$\sigma$ confidence region from measurements in the calibration sample. 
Circles with error bars represent measurements for individual galaxies adopting $R_{int}$ = 0.377.\label{fig:PAHratioSsil+models}}
\end{figure}

In the case of the screen model, we show in Appendix \ref{appendixb} that $\xi$ is a constant independent on \ssilu, which only depends on the opacities (relative to 9.8 \uu) at the wavelengths of the PAH bands (12.7 and 11.2 \uu) and the anchor points (5.5 and 14 \uu). Table \ref{table:diffopacity} gives the corresponding values for each extinction law.
The value of $\xi$ for the mixed geometry varies slightly with \ssilu. It takes increasingly negative values for more negative \ssilu, but converges to the value for the screen model at \ssil = 0.
This is shown in Figure \ref{fig:PAHratioSsil+models}, which compares the predicted relations between the ratio $R_{obs}$/$R_{int}$ and \ssil for the four extinction laws. 

The relation obtained empirically in our calibration sample (shaded grey area in Figure \ref{fig:PAHratioSsil+models}) is steeper than predicted for the screen geometry on all four extinction laws. Only the GC law falls within our 1-$\sigma$ confidence range. The relations are steeper for the mixed geometry, although the divergence becomes significant only at high opacity (\ssil $\lesssim$ -0.6). In this range the GC law becomes clearly too step, while the Milky Way law for R$_V$ = 5.5 crosses the empirical relation. The other two laws remain well below the empirical trend in the whole range.

One peculiarity of the mixed geometry model is the saturation of both \ssil and $R_{obs}$, which at high opacity converge to a maximum value (dependent on the extinction law) that cannot be surpassed regardless of the optical depth. We mark these limit values with
a solid star in Figure \ref{fig:PAHratioSsil+models}. For \ssil they range from -0.85 (diffuse ISM) to -1.39 (MW R$_V$=3.1), while maximum values for $R_{obs}$/$R_{int}$ are between 1.38 (diffuse ISM) and 1.91 (GC).

While our calibration sample contains only moderately obscured sources (\ssilu$\gtrsim$-1.1, $R_{obs}$/$R_{int}$$\lesssim$1.7), the FIR sample contains some sources with extreme ratios (up to $R_{obs}$/$R_{int}$ $\sim$ 4.5).
Such high ratios cannot be reproduced in a pure mixed geometry and require significant foreground extinction, as we show in the next section.

\subsection{Total extinction at 11.2 \um}

\begin{figure} 
\begin{center}
\includegraphics[width=8.4cm]{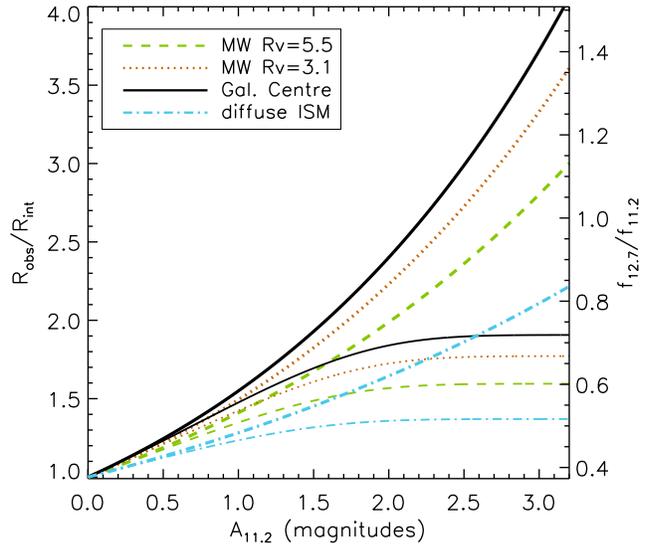}
\end{center}
\caption[]{Relation between the observed and intrinsic values of the 12.7/11.2 PAH flux ratio as a function of the total extinction at 11.2 \um for the extinction laws in Figure \ref{fig:extlaws}. Thick and thin lines correspond to screen and mixed dust geometries, respectively. The right axis indicates observed values of \pahratio{} assuming an intrinsic value $R_{int}$ = 0.377.\label{fig:attenuation}}
\end{figure}
 
The relation between $R_{obs}$/$R_{int}$ and the total extinction affecting the 11.2 \um PAH band, $A_{11.2}$, is derived in Appendix \ref{appendixa}.
For a given dust geometry the relation between $A_{11.2}$ and $R_{obs}$/$R_{int}$ is completely determined by the parameter $\alpha$ = \od{12.7}/\od{11.2}, which is a constant characteristic of each extinction law (see Table \ref{table:diffopacity}).

Figure \ref{fig:attenuation} shows that for a screen geometry $R_{obs}$/$R_{int}$ grows exponentially as a function of $A_{11.2}$, while tracks for the mixed geometry models converge to a constant value $R_{obs}$/$R_{int}$ = 1/$\alpha$ at very high optical depths.  
The actual value of the total extinction affecting the 11.2 \um PAH band in a given galaxy is constrained by the lower and upper limits defined by the tracks for the screen and mixed models, respectively. Because the tracks for the two geometries only diverge at $A_{11.2}$$\gtrsim$1 and $R_{obs}$ saturates by $A_{11.2}$$\sim$2 in the mixed model, we consider the screen model to be the best choice for estimating $A_{11.2}$ from $R_{obs}$ in all galaxies. This also prevents us from overestimating the total extinction of the 11.2 \um PAH band. 

Since the GC law with the screen model is the one that best approximates our empirical value for $\xi$, we calculate the total extinction affecting the 11.2 \um PAH band assuming that the value of $\alpha$ that applies to star-forming galaxies is also the one from the GC law. Table \ref{table:final-extinction} gives the values of $A_{11.2}$ and the correction factors needed to convert observed 11.2 \um PAH luminosities to intrinsic values, as a function of the observed 12.7/11.2 PAH flux ratio. 

We warn that the extinction correction will be biased if the true $\alpha$ differs from the value 0.525 characteristic of the GC law. We find some tentative evidence for a lower value $\alpha$ $\sim$ 0.4 in \S\ref{sec:applyfirsample}. A reliable empirical determination of $\alpha$ would require an independent measurement of the total extinction affecting either the 11.2 \um or 12.7 \um PAH bands, and is beyond the scope of this work.

\section{Application to the FIR sample}\label{sec:applyfirsample}

\begin{figure} 
\begin{center}
\includegraphics[width=8.4cm]{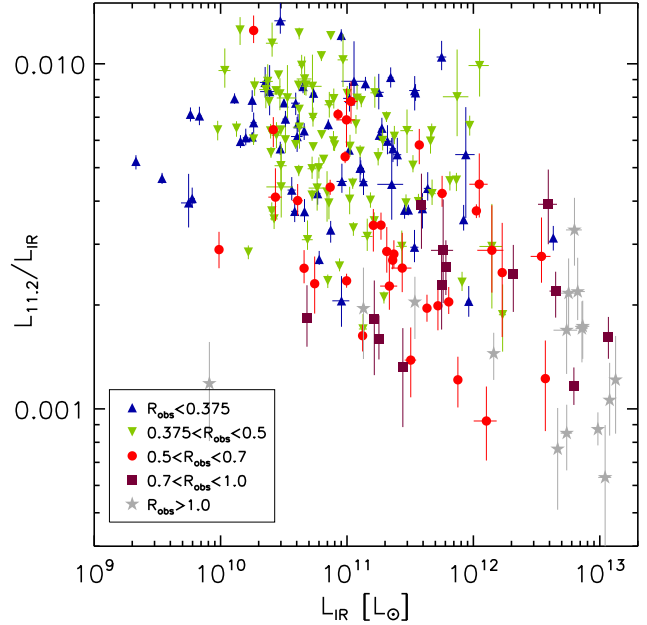}
\end{center}
\caption[]{Fractional contribution of the observed 11.2 \um PAH luminosity to the total IR [8--1000 \uu] luminosity (\lum{IR}) as a function of \lum{IR}{} for the FIR sample, with colour coding for the observed \pahratio.\label{fig:L112LIRratio}}
\end{figure} 

In this section we use a sample of galaxies with accurate measurements of the total (8--1000 \uu) infrared luminosity (\lum{IR}; see \S\ref{sec:firsample}) to check how the extinction affecting the 11.2 \um PAH band influences the relation between \lum{11.2}{} and \lum{IR}. 

Figure \ref{fig:L112LIRratio} shows the distribution of  \lum{11.2}/\lum{IR} as a function of \lum{IR} for individual galaxies in the sample. There is a clear trend towards lower \lum{11.2}/\lum{IR} at high \lum{IR} in spite of the high dispersion. 
In particular, sources in the ULIRG range (\lum{IR}$>$10$^{12}$ L$_\odot$) consistently have lower \lum{11.2}/\lum{IR} compared to less luminous galaxies, in agreement with previous findings in local ULIRG samples \citep[e.g.][]{Rigby08,Rieke09}. 
The colour coding in Figure \ref{fig:L112LIRratio} also shows an increase in the typical value of $R_{obs}$ from the top left to the bottom right. This suggests that higher extinction of the PAH emission contributes to the reduction in \lum{11.2}/\lum{IR} at high \lum{IR}.

\begin{figure} 
\begin{center}
\includegraphics[width=8.4cm]{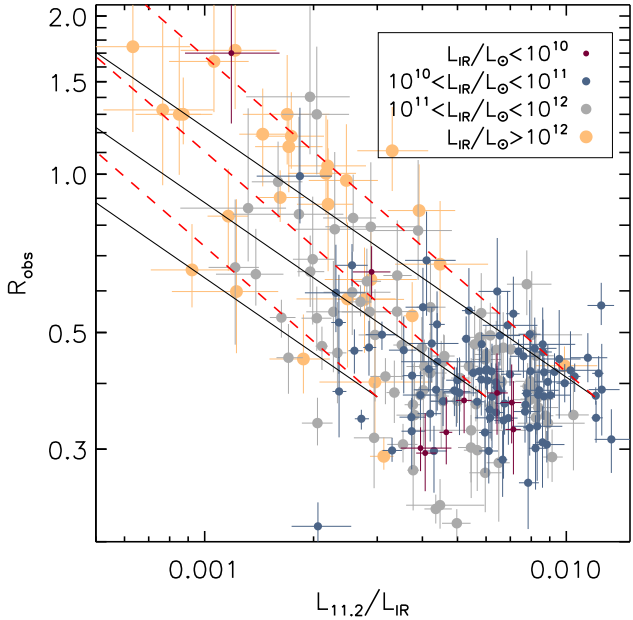}
\end{center}
\caption[]{Relation between the observed intensity ratio of the 12.7 and 11.2 \um PAH bands, $R_{obs}$ = \pahratio, and the fractional contribution of the 11.2 \um PAH luminosity to the total IR luminosity, \lum{11.2}/\lum{IR}, in the FIR sample. Symbol sizes indicate the range of \lum{IR}. Solid lines represent the expected relation for a screen geometry and an opacity ratio $\alpha$ = \od{12.7}/\od{11.2} = 0.525 (GC extinction law), for arbitrary values of the intrinsic \lum{11.2}/\lum{IR} = 0.003, 0.006, and 0.012. Dashed lines represent the same relations for $\alpha$ = 0.4.\label{fig:L112LIR-PAHratio}}
\end{figure}

Figure \ref{fig:L112LIR-PAHratio} shows $R_{obs}$ versus \lum{11.2}/\lum{IR} for the FIR sample. Most of the galaxies with \lum{IR}$<$10$^{11}$ L$_\odot$ concentrate around $R_{obs}$$\sim$0.4, consistent with low extinction. The range in \lum{11.2}/\lum{IR} for these galaxies is almost 1 dex, implying substantial dispersion in the ratio of the intrinsic \lum{11.2} to \lum{IR}, with no clear dependence on \lum{IR}. Factors other than extinction that may be responsible for this dispersion are a reduction in the PAH intensity relative to the MIR continuum at low metallicity \citep[e.g.][]{Wu06}, a reduction in the area of PDRs per unit SFR at higher SF density \citep{Elbaz11,Elbaz18,Magdis13}, and an increase in \lum{IR} for sources hosting an AGN. The contribution to \lum{IR} from dust-reprocessed AGN emission is small in most local LIRGs, but increases (on average) with \lum{IR} \citep[e.g.][]{Nardini10,Alonso-Herrero12}.

For ULIRGs there is a clear anti-correlation between \lum{11.2}/\lum{IR} and $R_{obs}$, although with high dispersion. Interestingly, lower luminosity galaxies with low \lum{11.2}/\lum{IR} also have unusually high $R_{obs}$, which places them in the same region as ULIRGs. This suggests that a common mechanism explains the anti-correlation between \lum{11.2}/\lum{IR} and $R_{obs}$ at all luminosities.
To check whether extinction can be such mechanism, we show in Figure \ref{fig:L112LIR-PAHratio} tracks for the expected relation between \lum{11.2}/\lum{IR} and $R_{obs}$ calculated for a screen model.
These tracks assume that the extinction affecting the PAH emission does not have a noticeable impact on \lum{IR}, and adopt values for the intrinsic \lum{11.2}/\lum{IR} of 0.003, 0.006, and 0.012. Changing this ratio displaces the tracks horizontally, but the slope is not affected since it only depends on $\alpha$ = \od{12.7}/\od{11.2} (see \S\ref{appendixb}).

The slope predicted for the GC law ($\alpha$ = 0.525; solid lines) is roughly consistent with the data, given the high dispersion, although a slightly lower value of $\alpha$ $\sim$ 0.4 is preferred (dashed lines). This suggests that extinction is indeed the dominant factor driving the correlation between $R_{obs}$ and \lum{11.2}/\lum{IR}, and an important contributor to the lower \lum{11.2}/\lum{IR} in ULIRGs. We note that alternative interpretations, such as destruction of PAH molecules by stronger radiation fields, reduction of the relative volume of PDRs in regions of high-density star formation, or higher \lum{IR} in AGN hosts cannot account for the increase in $R_{obs}$ at low \lum{11.2}/\lum{IR}.

\begin{figure} 
\begin{center}
\includegraphics[width=8.4cm]{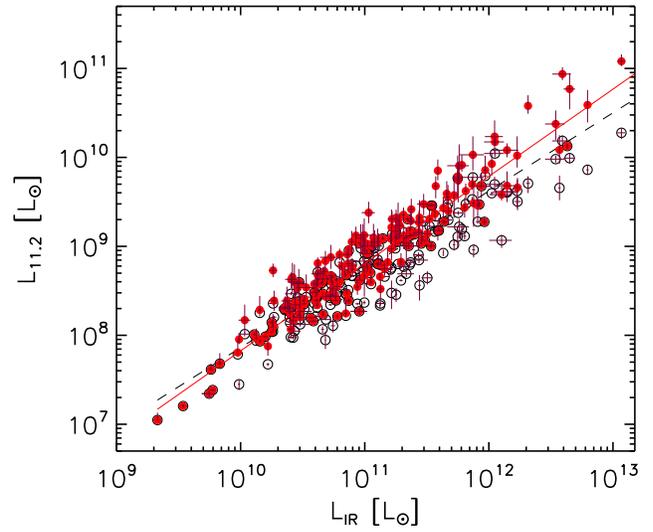}
\end{center}
\caption[]{11.2 \um PAH luminosity as a function of the total IR luminosity in the FIR sample. Open symbols represent 11.2 \um luminosities uncorrected for extinction, while solid ones include the extinction correction determined from their observed \pahratio{} using our empirical calibration. The solid and dashed lines represent the best power-law fits with and without extinction correction, respectively.\label{fig:L112-LIR}}
\end{figure}

We use our calibration of the intrinsic to observed 11.2 \um PAH luminosity ratio as a function of $R_{obs}$ (Table \ref{table:final-extinction}) to compute intrinsic 11.2 \um PAH luminosities for the FIR sample. For sources with $R_{obs}$ $<$ 0.377 we assume no extinction. 
Figure \ref{fig:L112-LIR} shows the relation between \lum{11.2}{} and \lum{IR}{} before (open symbols) and after (solid symbols) applying the extinction correction to \lum{11.2}. A least squares fit to a power-law relation: \lum{11.2}{} $\propto$ \lum{IR}$^{\beta}$ is shown as dashed and solid lines, respectively.
 
The extinction correction decreases the dispersion in the correlation between \lum{11.2}{} and \lum{IR}{} only marginally, from 0.211 dex to 0.209 dex. However, given the large uncertainties in the measurements of $R_{obs}$ and the corresponding error in the total extinction, the intrinsic dispersion may be significantly lower. 
The slope of the best-fitting power-law relation is $\beta$=0.882$\pm$0.020 before and $\beta$=0.985$\pm$0.033 after correcting for extinction. This suggests that a constant value of \lum{11.2}/\lum{IR}{} is maintained through almost 4 orders of magnitude in \lum{IR}, although the number of sources outside the 10$^{10}$$<$\lum{IR}/L$_\odot$$<$10$^{12}$ range is too small to be conclusive.
If confirmed in a larger sample, this would eliminate the need to invoke PAH suppression to explain the apparent low \lum{11.2}/\lum{IR}{} in local ULIRGs, which would be just a consequence of the increased frequency of high column densities at higher \lum{IR}. This is in agreement with the finding by \citet{Rieke09} in local star-forming galaxies of a reduction of the 24 \um continuum luminosity relative to \lum{IR} at \lum{IR}{} $>$ 10$^{11}$ L$_\odot$, which they attribute to higher opacity. 

\section{Prospects for JWST and SPICA}

The Mid-InfraRed Instrument \citep[MIRI;][]{Wright15} onboard \textit{JWST} will provide integral field spectroscopy of the 11.2 \um and 12.7 \um PAH bands for galaxies up to $z$$\sim$1. The spectral resolution and sensitivity will improve by one and two orders of magnitude, respectively, compared to \textit{Spitzer}/IRS. This will allow for more accurate determinations of PAH luminosities as well as better separation of the multiple components in the PAH bands. 

While the 11.2 \um PAH band shifts out of the MIRI range at $z$$\sim$1.5, the 6.2 \um and 3.3 \um bands are observable up to $z$$\sim$3.5 and $z$$\sim$7.7, respectively. The 3.3 \um band seems particularly promising since it shares a common origin with the 11.2 \um band \citep[e.g.][]{vanDiedenhoven04} and comparable levels of extinction, and will also be observable in the local Universe with \textit{JWST}/NIRSpec.

At the same time, the SPace Infrared-telescope for Cosmology and
Astrophysics \citep[SPICA;][]{Roelfsema18}, could offer a high
spectroscopic sensitivity and continuous coverage in the wavelength
range between 12–-230 \uu, covering the MIR PAH bands all the way to the
highest redshifts \citep[][, Spinoglio et al. submitted]{Kaneda17}.

Improved measurements of multiple PAH ratios, also including spatial information in the case of relatively nearby galaxies (the PSF FWHM of MIRI is $\sim$0.3'' at 8 \uu) has the potential to reveal a wealth of information about the physical conditions in the PDRs, that is complementary to that obtained from fine structure and $H_2$ lines. 

It is likely that besides the 11.2 \um and 12.7 \um PAH bands, other bands or band components also have constant intrinsic ratios under some circumstances, which could be used to estimate the extinction affecting the PAH emission with a method comparable to the one presented in this work. 
Estimating the extinction on the individual PAH bands not only removes a source of uncertainty in the determination of other physical parameters, but also constrains the shape of the extinction law, which in turn informs about the chemical and mineralogical composition of the obscuring dust.

The recent discovery of dusty hyper-luminous galaxies at $z$$\sim$5--6 \citep[``870 \um risers'';][]{Riechers17, Yan19} shows that PAH emission is likely to be detectable with \textit{JWST} and \textit{SPICA} into the Reionisation epoch.
 A better understanding of PAH emission and mid-IR extinction is therefore essential to trace the star formation in dusty galaxies across cosmic time.

\section{Conclusions}

In this work we have shown that on galaxy-wide scales extinction is the main driver of variation in the intensity ratio between the 12.7 \um and 11.2 \um PAH bands. The ratio of their equivalent widths (which is insensitive to extinction) indicates that, after accounting for the photometric uncertainty, the intrinsic dispersion in \pahratio{} is $\lesssim$5\%. 
This provides a new method for estimating the total extinction affecting the 11.2 \um PAH band based on the relation between the observed ($R_{obs}$) and intrinsic ($R_{int}$) values of \pahratio. 
Unlike other methods based on hydrogen recombination or molecular line ratios, $R_{obs}$ directly traces the extinction affecting the PAH emission, and therefore provides unbiased estimates of the intrinsic PAH luminosity in galaxies with substantial obscuration. Furthermore, this method is also useful for composite sources, since $R_{obs}$ is not affected by the emission of the AGN (unlike the recombination line ratios and \ssilu). 

In galaxies that do not host AGN, both $R_{obs}$ and \ssil trace the extinction towards the star-forming regions. This allows to calibrate the relation between $R_{obs}$ and the total extinction affecting the 11.2 \um PAH band and to put constraints on the mid-IR extinction law for star-forming regions. The relation between $R_{obs}$ and \ssil is given by: $\ln$($R_{obs}$) = $\ln$($R_{int}$) + $\xi$\ssilu, with $R_{int}$ = 0.377 $\pm$ 0.003 (stat) $\pm$ 0.016 (sys) and $\xi$ = -0.435 $\pm$ 0.013 providing the best fit for our calibration sample.
Comparison with four widely used extinction laws derived from lines of sight within the Milky Way shows that only the Galactic Center (GC) extinction law is consistent with the empirical $R_{obs}$-\ssil relation at the 1-$\sigma$ level, while all the others clearly under-predict $R_{obs}$ for a given \ssilu.

We show that the relation between $R_{obs}$ and the total extinction in the 11.2 \um PAH band, $A_{11.2}$, is roughly linear and independent on the geometry of the dust distribution for $A_{11.2}$$\lesssim$1. However, $R_{obs}$ grows exponentially at $A_{11.2}>$1 for the screen geometry, while for the mixed geometry it converges to a constant value. Accordingly, only dust in the foreground of the PAH-emitting regions can reproduce the very high $R_{obs}$ found in some galaxies, as is also the case for deep silicate absorption.

In a sample of 215 normal and active galaxies with accurate measurements of the total infrared luminosity (\lum{IR}), the ratio \lum{11.2}/\lum{IR}{} decreases at high \lum{IR}, in agreement with previous results on local ULIRGs. 
There is a correlation, although with high dispersion, between \lum{11.2}/\lum{IR}{} and $R_{obs}$ that is independent on \lum{IR}, which suggests that the main reason for the decrease in \lum{11.2}/\lum{IR}{} is higher average extinction at high \lum{IR}. Alternative interpretations, such as a reduction in PDR volume or PAH destruction in environments with high star formation density, do not predict the correlation with $R_{obs}$. The slope of the dependence of $R_{obs}$ with \lum{11.2}/\lum{IR}{} seems steeper than predicted for the GC law, pointing to a smaller value for the opacity ratio \od{12.7}/\od{11.2} $\sim$ 0.4. However, higher S/N measurements of the PAH bands in a sample with independent estimates of the total extinction are required for confirmation.

The relation between \lum{11.2} and \lum{IR}{} is reproduced by a power-law: \lum{11.2} $\propto$ \lum{IR}$^\beta$. For the observed value of \lum{11.2}, the best fit is obtained for $\beta$=0.882$\pm$0.020. After correcting for extinction, it increases to $\beta$=0.985$\pm$0.033.
Therefore, the intrinsic \lum{11.2} is roughly proportional to \lum{IR}{} over almost 4 orders of magnitude (\lum{IR}{} = 10$^9$--10$^{13}$ L$_\odot$).

These results consolidate the 11.2 \um PAH band as a robust tracer of star formation in galaxies, while at the same time manifest the biases that may occur if the extinction is not properly taken into account. 

\section*{Acknowledgements}

We thank the anonymous referee for useful comments and suggestions that helped improve the paper.
This work is based on observations made with the \textit{Spitzer Space
Telescope}, which is operated by the Jet Propulsion Laboratory, Caltech
under NASA contract 1407.
This research has made use of the NASA/IPAC Extragalactic Database (NED) which is operated by the Jet Propulsion Laboratory, California Institute of Technology, under contract with the National Aeronautics and Space Administration. 
A.H.-C. acknowledges funding by the Spanish Ministry of Science, Innovation and Universities under grants AYA2015-63650-P and ESP2017-83197-P. 
A.A.-H. acknowledges support from grant PGC2018-094671-B-I00 (MCIU/AEI/FEDER,UE). A.A.-H., P.P.-G. M.P.-S., S.A., A.L. and J.P. work was done under project  No. MDM-2017-0737 Unidad de Excelencia ``Mar\'ia de Maeztu'' - Centro de Astrobiolog\'ia (INTA-CSIC).
A.L. acknowledges the support from Comunidad de Madrid through the Atracci\'on de Talento Investigador Grant 2017-T1/TIC-5213.
G.E.M. acknowledges the Villum Fonden research grant 13160 ``Gas to stars, stars to dust: tracing star formation across cosmic time'' and the Cosmic Dawn Center of Excellence funded by the Danish National Research Foun-dation under then grant No. 140.
M.P.S. acknowledges support from the Comunidad de Madrid, Spain, through the Atracci\'on de Talento Investigador Grant 2018-T1/TIC-11035.
J.P.L. acknowledges financial support by the Spanish MICINN under grant AYA2017-85170-R.

\section*{Data Availability}

The data underlying this article will be shared on reasonable request to the corresponding author.

\onecolumn

\begin{deluxetable}{l cccccccc}
\tabletypesize{\scriptsize}
\tablewidth{0pc}
\tablecolumns{9}
\tablecaption{Extinction law dependent quantities\label{table:diffopacity}}
\tablehead{\colhead{Extinction law} & \colhead{\od{5.5}\tablenotemark{*}} & \colhead{\od{11.2}\tablenotemark{*}} & \colhead{\od{12.7}\tablenotemark{*}} & \colhead{\od{14.0}\tablenotemark{*}} & \colhead{$\alpha$\tablenotemark{a}} & \colhead{$k$\tablenotemark{b}} & \colhead{$\Delta\tau$\tablenotemark{c}} & \colhead{$\xi$\tablenotemark{d}}}
\startdata
 MW Rv=5.5 (Weingartner \& Draine 2001) &    0.438 &    0.662 &    0.415 &    0.313 &    0.627 &    0.639 &   -0.247 &   -0.386\\
               MW Rv=3.1 (Draine 2003) &    0.261 &    0.623 &    0.352 &    0.241 &    0.565 &    0.751 &   -0.272 &   -0.362\\
             GC (Chiar \& Tielens 2006) &    0.299 &    0.617 &    0.323 &    0.315 &    0.525 &    0.692 &   -0.293 &   -0.424\\
        diffuse ISM (Shao et al. 2018) &    0.418 &    0.701 &    0.512 &    0.434 &    0.730 &    0.572 &   -0.189 &   -0.331\\    
\enddata
\tablenotetext{*}{normalized to \od{9.8} = 1}
\tablenotetext{a}{$\alpha$ = \od{12.7}/\od{11.2}}
\tablenotetext{b}{$k$ = -\ssilu/\od{9.8}}
\tablenotetext{c}{$\Delta\tau$ = (\od{12.7} - \od{11.2})/\od{9.8}}
\tablenotetext{d}{$\xi$ = $\Delta\tau$/$k$}
\end{deluxetable}

\begin{deluxetable}{ccc}
\tabletypesize{\scriptsize}
\tablewidth{0pc}
\tablecolumns{3}
\tablecaption{Calibration for total extinction\label{table:final-extinction}}
\tablehead{\colhead{\pahratio\tablenotemark{a}} & \colhead{$A_{11.2}$\tablenotemark{b}} & \colhead{$L_{\rm{11.2}}^{int}$/$L_{\rm{11.2}}^{obs}$\tablenotemark{c}}} \startdata
0.377  &    0.000  &    1.000 \\
0.400  &    0.135  &    1.133 \\ 
0.425  &    0.274  &    1.287 \\
0.450  &    0.404  &    1.451 \\
0.475  &    0.528  &    1.626 \\
0.500  &    0.645  &    1.811 \\
0.525  &    0.756  &    2.007 \\
0.550  &    0.862  &    2.213 \\
0.575  &    0.964  &    2.430 \\
0.600  &    1.061  &    2.657 \\
\enddata
\tablenotetext{a}{Observed flux ratio between PAH bands at 12.7 and 11.2 \uu.}
\tablenotetext{b}{Total extinction (in magnitudes) affecting the 11.2 \um PAH band.}
\tablenotetext{c}{Ratio between the intrinsic and observed luminosity of the 11.2 \um PAH band.}
\tablenotetext{Note.}{This table is available in its entirety in a machine-readable form in the online version of the article and at http://www.denebola.org/ahc/PAHextinction/extcal\_PAH112.txt}
\end{deluxetable}

\appendix

\section{Derivation of relations between \pahratio{} and $A_{11.2}$}\label{appendixa}

The geometric distribution of the PAH emitting molecules and the obscuring dust determines how the ratio between the observed and intrinsic flux decreases as a function of the optical depth. Two ideal cases are often considered: the screen model and the fully mixed model. In the screen model all the obscuring dust is located in the foreground, between the PAH emitting regions and the observer, while in the fully mixed model all the dust is located in the PAH emitting regions, homogeneously distributed among the stars. The corresponding relations are \citep{Smith07}: 
\begin{subequations}\label{eqn:fobsfint}
\begin{align}
 \frac{f^{obs}(\lambda)}{f^{int}(\lambda)} & = e^{-\tau(\lambda)} & & (screen \, model) \\
 \frac{f^{obs}(\lambda)}{f^{int}(\lambda)} & = \frac{1 - e^{-\tau(\lambda)}}{\tau(\lambda)} & & (mixed \, model)
\end{align}
\end{subequations}

\noindent where $f^{obs}(\lambda)$ and $f^{int}(\lambda)$ are, respectively, the observed and intrinsic flux density of the source at wavelength $\lambda$, and $\tau$($\lambda$) is the optical depth. 
From Equations (\ref{eqn:fobsfint}) we get that the observed ratio between the fluxes of the 12.7 and 11.2 \um PAH bands, $R_{obs}$, is related to its intrinsic value, $R_{int}$, and the optical depths at 11.2 \um and 12.7 \um by:
\begin{subequations}\label{eqn:robsrint}
\begin{align}
R_{obs} & = R_{int} \, e^{\tau(11.2)-\tau(12.7)} & & (screen) \\
R_{obs} & = R_{int} \, \frac{1-e^{-\tau(12.7)}}{1-e^{-\tau(11.2)}} \frac{\tau(11.2)}{\tau(12.7)}                               & & (mixed)
\end{align}
\end{subequations}

\noindent For a given extinction law, the ratio $\alpha$ = $\tau$(12.7)/$\tau$(11.2) is constant. Therefore Eq. (\ref{eqn:robsrint}) can be rewritten as:

\begin{subequations}\label{eqn:robsrint3}
\begin{align}
R_{obs} & = R_{int} \, e^{(1-\alpha)\tau(11.2)} & & (screen) \\
R_{obs} & = R_{int} \, \frac{1-e^{-\alpha\tau(11.2)}}{1-e^{-\tau(11.2)}} \frac{1}{\alpha} & & (mixed)
\end{align}
\end{subequations}

\noindent The total extinction (sometimes also known as attenuation) at the wavelength $\lambda$, $A$($\lambda$), is defined as: \linebreak $A$($\lambda$) = -2.5 $\log_{10}$($f^{obs}$($\lambda$)/$f^{int}$($\lambda)$). By substituting $\lambda$ = 11.2 \um in Equations \ref{eqn:fobsfint} we obtain:
\begin{subequations}\label{eqn:attenuation}
\begin{align}
A(11.2) & = \frac{2.5}{\ln 10} \tau(11.2) & & (screen) \\
A(11.2) & = \frac{2.5}{\ln 10} \ln \Big{(}\frac{\tau(11.2)}{1 - e^{-\tau(11.2)}}\Big{)} & & (mixed)
\end{align}
\end{subequations}

\noindent which together with Equations (\ref{eqn:robsrint3}) give a parametric relation between $A$(11.2) and $R_{obs}$.\\

\section{Derivation of relations between \ssilu, $\tau$(9.8), and $R_{obs}$/$R_{int}$}\label{appendixb}

The silicate strength is defined \citep[e.g.][]{Spoon07} as: \ssil = $\ln$ [$f^{obs}$($\lambda_p$)/$f^{cont}$($\lambda_p$)], where $\lambda_p$ is the wavelength of the peak of the silicate feature (usually $\lambda_p$$\sim$9.8 \um when found in absorption) and $f^{cont}$($\lambda_p$) is the continuum that would be measured at $\lambda_p$ in absence of silicate emission/absorption.
When the silicate absorption is produced by cold dust (i.e., no radiative transfer effects), \ssil traces the optical depth at 9.8 \uu, $\tau$(9.8).

To estimate $f^{cont}$($\lambda_p$), we use a power-law interpolation
between anchor points at $\lambda_1$ = 5.5 \um and $\lambda_2$ = 14 \uu, where the dust opacity is $\sim$2--5 times lower compared to $\lambda_p$, depending on the extinction law (see Figure \ref{fig:extlaws}).
A realistic determination of $\tau$(9.8) from \ssil requires to also take into account the opacity at the anchor points $\lambda_1$ and $\lambda_2$. 
In the following we derive the relation for the screen model. The relation for the mixed model can be obtained in the same way.\\ 

Lets assume that the intrinsic continuum between $\lambda_1$ and $\lambda_2$ is a power-law with spectral index $\alpha_{int}$. Then the intrinsic fluxes of the spectrum at $\lambda_1$ and $\lambda_2$ are related by:

\begin{equation}\label{eqn:fintalpha}
f^{int}(\lambda_1) = f^{int}(\lambda_2) \Big{[}\frac{\lambda_1}{\lambda_2}\Big{]}^{\alpha_{int}}
\end{equation}

\noindent Using Equation \ref{eqn:fobsfint} for the screen model, the relation between observed fluxes is then:

\begin{equation}\label{eqn:fobsalpha}
f^{obs}(\lambda_1) = f^{obs}(\lambda_2) \Big{[}\frac{\lambda_1}{\lambda_2}\Big{]}^{\alpha_{int}} e^{\tau(\lambda_2) - \tau(\lambda_1)}
\end{equation}

\noindent and the spectral index of the interpolated continuum between the observed anchor points is then: 

\begin{equation}\label{eqn:alphaobs}
\alpha_{obs} = \frac{\ln \frac{f^{obs}(\lambda_1)}{f^{obs}(\lambda_2)}}{\ln \frac{\lambda_1}{\lambda_2}} = \alpha_{int} + \frac{\tau(\lambda_2) - \tau(\lambda_1)}{\ln \frac{\lambda_1}{\lambda_2}}
\end{equation}

\noindent Therefore, the interpolated continuum at the peak wavelength of the silicate absorption, $\lambda_p$, is:
\begin{equation}\label{eqn:fcont}
f^{cont}(\lambda_p) = f^{obs}(\lambda_1)\Big{[}\frac{\lambda_p}{\lambda_1}\Big{]}^{\alpha_{obs}} = f^{int}(\lambda_p) e^{-\tau(\lambda_1)} \Big{[}\frac{\lambda_p}{\lambda_1}\Big{]}^{\alpha_{obs} - \alpha_{int}}
\end{equation}

\noindent Substituting in the expression for \ssil above, after some simplification we arrive at:
\begin{equation}\label{eqn:ssildef}
S_{sil} = -\tau(\lambda_p) + \tau(\lambda_1) + \eta[\tau(\lambda_1) - \tau(\lambda_2)],\hspace{0.8cm} \eta = \frac{\ln \lambda_p - \ln \lambda_1}{\ln \lambda_1 - \ln \lambda_2}
\end{equation}

\noindent where the constant $\eta$ only depends on the wavelengths of the anchor points and takes the value $\eta$ = -0.618 for $\lambda_1$ = 5.5 \uu, $\lambda_2$ = 14 \uu, and $\lambda_p$ = 9.8 \uu. Taking $\lambda_p$ = 9.8 \um and reorganising:

\begin{equation}\label{eqn:ssilsimple}
S_{sil} = -k \tau(9.8), \hspace{0.8cm} k = 1 - \frac{\tau(\lambda_1)}{\tau(9.8)} - \eta\frac{\tau(\lambda_1) - \tau(\lambda_2)}{\tau(9.8)}
\end{equation}

\noindent where the constant $k$ depends on the extinction law (and the anchor points via $\eta$).\\

\noindent Now we can also rewrite Equation \ref{eqn:robsrint} for the screen model as a function of $\tau$(9.8):

\begin{equation}\label{eqn:logrobs}
\ln R_{obs} = \ln R_{int} + \tau(9.8)\Delta\tau, \hspace{0.8cm} \Delta\tau = \frac{\tau(12.7) - \tau(11.2)}{\tau(9.8)}
\end{equation}

\noindent where $\Delta\tau$ is also a constant that depends only on the extinction law. Finally, substituting $\tau$(9.8) = -\ssilu/$k$, we arrive at:

\begin{equation}\label{eqn:logrelation}
\ln R_{obs} = \ln R_{int} + \xi S_{sil}, \hspace{0.8cm} \xi = \frac{\Delta\tau}{k}
\end{equation}

For the mixed model, $R_{obs}$ cannot be expressed explicitly as a function of \ssilu, but the relation between the two is given by the parametric equations:

\begin{subequations}\label{eqn:parametric}
\begin{align}
S_{sil} & = \ln\frac{x(9.8; t)}{x(\lambda_1; t)} + \eta \ln \frac{x(\lambda_2; t)}{x(\lambda_1; t)} \\
R_{obs} & = R_{int} \frac{x(12.7; t)}{x(11.2; t)}\\ 
x(\lambda; t) & = \frac{1 - e^{-t \frac{\tau(\lambda)}{\tau(9.8)}}}{t \frac{\tau(\lambda)}{\tau(9.8)}} 
\end{align}
\end{subequations}

\noindent where the parameter $t$ $\equiv$ $\tau$(9.8) is the optical depth at the peak of the silicate feature.

\end{document}